\documentclass[manuscript]{aastex}

\newcommand{\be}{\begin{equation}}
\newcommand{\ee}{\end{equation}}
\newcommand{\bea}{\begin{eqnarray}}
\newcommand{\eea}{\end{eqnarray}}

\shorttitle{Preferred frame}
\shortauthors{Henriksen}
\begin{document}

\title{Cosmology without an Initial Singularity}

\author{R.N. Henriksen}

\affil {Queen's University at Kingston, Ontario, Canada}

\email{henrikr@queensu.ca}

\begin{abstract}
{We present a model of the expanding Universe that begins in a zero mass/energy  vacuum state.  The Universe results from the  spontaneous  breaking  of the electroweak symmetry, wherein the vacuum  with positive expectation energy (i.e. `dark energy' ) produces the dark and visible matter and launches the expansion.  A Hubble-Lem\^aitre constant of $\sim 73.36 km/sec/Mpc$ and the equipartition red shift  are part of the model.  The current value of Einsein's $\Lambda\sim 1.29\times 10^{-56} cm^{-2}$. We require the age of the Universe to be $\sim 28$ billion years. Therefore one may find galaxies, quasars and  black holes at higher redshifts than the maximum value now known, but otherwise the description of the development of structure is not greatly changed. A Dimensional argument based on the Buckingham theorem applied to the Universe, is given in an appendix to justify the dark energy density $\propto 1/t^2$.  Such behaviour produces rapid but not exponential Universal expansion. We speculate regarding the nature of each Universe energy component, due to a coincidence between the sine of the electroweak Weinberg angle and the ratio of matter to dark energy.  }  
\end{abstract}
\maketitle


\section{Introduction}

The current `concordance model' of cosmology has a string of predictive sucesses (e.g. \cite{OS1995}, \cite{Peter2013}, 
\cite{Planck2018}), with the anticipation of \cite{OS1995} regarding the  importance of the vacuum energy being remarkable.  The origin of the local gravitational structure in quantum fluctuations plus inflation, as observed in the CMB and in the present epoch; is well described pedagogically in \cite{Peter2013}.  Predictions regarding primordial gravitational waves  remain as outstanding tests. Any competing cosmology must do at least as well.

However the existence of the initial singularity remains a part of the model that defies our present physics. Moreover Roger Penrose (e.g. \cite{Pen2004}, \cite{Pen2010}) has persistently called attention to the cosmological problem presented by the second law of thermodynamics. Assuming the Bekenstein-Hawking entropy of a black hole, one requires the gravitational collapse degrees of freedom to  be minimized in order that the Universe begin in a low entropy state. Together with the second law this implies the normal arrow of time. This argues against the existence of  primordial super massive black holes for which there is no current evidence, although intermediate scale objects may be possible (e.g.\cite{BH1979}).  

Currently (e.g. \cite{Planck2018}) the equation of state of the dark energy  is consistent with the cosmological constant, although a time dependent scalar field (`Quintessence') with a dominant self-interacting potential has the same equation of state.  The vacuum term should always be Lorentz invariant, which requires the pressure to be the negative of the energy density. 



I wish to call attention in this article to an older and simpler idea regarding the dark energy (e.g. \cite{Hen1982}, \cite{OJ1987}, \cite{RP1988} ) that may describe the initial state of the Universe. This idea accepts that Einstein's cosmological term $\Lambda$, once translated to the energy momentum tensor part of the field equations, can be regarded as the vacuum term and need not be constant in time. The mechanism for the time dependence is left undetermined except for Dimensional (we use capital `D' to mean physical parameters and small `d' for geometrical dimensions).  The Dimensional argument is not entirely without physical content however, because of  relations  between Universe properties dictated by the Buckingham theorem.
Only the total mass/energy is conserved in such models, so that  the balance between matter and vacuum energy may evolve in time.

A `natural'  temporal variation is given Dimensionally as an inverse length squared . This is explicitly (\cite{Hen1982})  for the $\Lambda$ function
 \be
 \Lambda(t)=\frac{\lambda}{c^2t^2},\label{eq:lambda1}
 \ee
 where $t$ is cosmic time, $\lambda$ is a number and $c$ is the absolute speed of light. One should note that this implies that an associated scale $\ell=1\sqrt{\Lambda}$ is essentially equal to the apparent horizon at each epoch. The scale $\ell$ is also roughly the scale at which the gravitational binding energy of the matter is about equal to  the vacuum energy.

A  temporal variation  of the vacuum energy has been discussed elsewhere, also before the arrival of extensive cata on the CMB. Ratra and Peebles (\cite{RP1988}) studied it as an early form of the inflaton field and Olson and Jordan (\cite{OJ1987}) discussed the implied age of the Universe in terms of various power laws. The latter paper gives many earlier references \footnote{The reference to Henriksen 1982 was omitted due probably to appearing in an European journal} in which physical  justifications are attempted. Ultimately the inflaton field and the power laws remain essentially ad hoc, except for Dimensional analysis and the Dirac argument that we use.

The  observational inputs to our model amount to the ratio of all matter to dark energy, $\Omega_{mv}$, the red shift of equipartition (\citet{Planck2018}), and the temperature at the electroweak symmetry breaking $T\sim 160GeV$. We iterate on the initial ratio of matter to photon energy until the current scale factor agrees with an initial assumption, and we iterate on the $\lambda$ parameter until the current ratio of matter to dark energy is obtained. With these inputs and procedures; that age of the Universe, the current Hubble-Lemaitre `constant', and the current cosmological `constant' are predicted more or less uniquely.

The thermodynamical evolution of the model is similar to the concordance model, but there is no inflation (spatial flatness is here ensured by the initial vacuum state) and the surface of last scattering is at a much later epoch than currently believed (assuming the same red shift for the CMB) . 

 

\section{Physical Description of the Model}

Our model (to be demonstrated below) is a Universe that begins in a unstable, static state of zero total energy.  At some epoch $t_1$ (which is subsequently adopted as the unit of time) the vacuum component spontaneously  breaks the temporal symmetry, creating a rapid  transition to a mixture of matter (including dark matter) and `dark energy'. This dynamic  state is one of  `a priori' spatially flat expansion  with $\Omega_m+\Omega_v=1$  . The ratio of matter density to dark energy can be fixed asymptotically to a value  $\Omega_{mv}\sim 0.46$ (from $\Omega_m=31.5 \pm1.3\%$ and $\Omega_m+\Omega_v=1$; \citet{AKW2020},\cite{Planck2018}), by adjusting one model constant that defines the post instability evolution dynamically. This  scenario follows  from  the equations of General Relativity, homogeneity, and assumption (\ref{eq:lambda1}); taken to hold backwards in time to the tr epoch of vacuum instability. The Universe may be in the zero energy state for an indefinite earlier time because of the temporal symmetry (i.e. static).

One can refer to this static, zero energy state as a `classical vacuum', but before $t_1$ we are beyond  classical validity and  we must necessarily hypothesize a quantum vacuum. This would be comprised of various particle fields as described by quantum field theory, with their characteristic fluctuations and expectation values.  A  detailed discussion is beyond the scope of this model, but this assumption does allow some useful speculations. 

The question arises as to how the temporal symmetry (static vacuum) is broken. This question is similar to determining the origin of the `big bang'. In our case this event is an example of spontaneous symmetry breaking of the metastable vacuum in the net zero energy Universe.  The vacuum symmetry breaking  produces matter with positive energy (exciting it from the negative energy state) and launches the expansion.  This suggests that the observable Universe expansion is a `pseudo' Goldstone boson (see \cite{LS2013} for an interesting qualitative presentation) at large scale. 

The  positive matter density must involve the gauge fields associated with the  initial symmetry.
We need these gauge fields to be spontaneously mixed in the presence of the Higgs field to yield massive  particle excitations according to the Brout, Englert, Higgs (BEH) mechanism. The descendants of these particles provide the total matter content of the Universe.

The matter gauge fields that seem to fit with the future history of the Universe (see below) are those associated with the electroweak gauge theory. At a temperature $T\sim 160 GeV$ the massless fields spontaneously mix to give the massive $Z^o$ vector boson, and the massive $W^{\pm}$ conjugate  vector bosons in the presence of the Higgs scalar boson. These massive vector particles decay quickly to hadrons and leptons and so begin the normal matter evolution to low temperature. The mass distribution  of these descendants are presumably due to interaction with the Higgs field, whose excitations also decay. There is a remarkable coincidence between $\Omega_{mv}$ and the Weinberg angle
defining the boson mixtures, that encourage us to speculate about the nature of the  matter and the dark energy. It may of course be simply a coincidence, but it could be relevant to  the necessary incomplete matter/antimatter annihilation.


It is possible that the vacuum state is not static and some quintessence type self-interacting scalar field is decaying and yielding the $1/t^2$ time dependence from the Planck epoch to the instability.  Perhaps it would begin only at a GUT (Grand Unified Theory) spontaneous symmetry breaking. This would have the advantage of connecting the earlier (Planck or GUT) value of $\Lambda$ to the current value. However this  simply displaces the problem in an `ad hoc' manner, and very little is certain about a hypothetical GUT.

Our model requires an older Universe (by a factor $2$), but the surface of last scattering is at the usual redshift $z_s\approx 1087$. 
There may consequently be time between equipartition  and  $z_s$ for  stronger perturbations  to grow  near the last scattering surface. These would presumably be adiabatic and at physical scales above the Jeans length. The linear analysis above the Jeans scale gives growth $\propto t^{0.57}$ with the parameters of our model.


  In \cite{Hen1982} the  usual FLRW formulation of cosmology was used. We revisit and expand this treatment in the next section  using the more intuitive Misner ( \cite{Mis1969} chapter 3) formulation of the field equations, assuming flat FLRW symmetry. The zero energy initial state was already found in \cite{Hen1982}, but not  analyzed in terms of modern cosmological parameters. 
  
In an appendix we give an argument based on the Buckingham theorem of Dimensional analysis (\cite{HIr2019}, see also \cite{Hen2015}) applied to the Universe. This suggests numbers of order one at our epoch may have always been of order one. If so, the inverse square temporal dependence of the vacuum energy ($\Lambda$) follows. This is much like the argument that Dirac used to justify a temporal variation in the gravitational constant.

\section{Basic Formulation}

We select the independent equations from (\cite{Mis1969}; see also \cite{HEW1983} and references therein) in the homogeneous form using co-moving coordinates $\{r,t\}$;
\bea
ds^2&=&dt^2-e^{\omega(t)} dr^2-R(r,t)^2(d\theta^2+(\sin{\theta})^2d\phi^2),\label{eq:metric}\\
\partial_r m(r,t)&=& 4\pi R^2\rho\partial_rR,\label{eq:mass}\\
\partial_t\omega&=& -4\frac{\partial_tR}{R}-2\frac{\partial_t\rho}{p+\rho},\label{eq:energycons}\\
e^{-\omega}(\partial_rR)^2&-&(\partial_tR)^2=1-\frac{2m(r,t)}{R},\label{eq:energyinit}\\
\partial_tm(r,t)&=&-4\pi R^2p\partial_tR.\label{eq:massevolution}
\eea
We have set $G=c=1$; $m(r,t)$ is the total mass-energy inside a co-moving spherical shell labelled $r$; $R$ is the circumferential radius; $\rho(t)$ is the density of mass-energy; $p(t)$ is the total pressure. The spherical angles $\theta$ and $\phi$ are constant (co-moving) in radial motion  and $ds$ is the metric distance. The last of these equations is normally satisfied identically when the other equations are satisfied, but it is necessary  when the scale dependence of the various individual components of $\rho$ are to be found.

Our working equations follow in our assumed symmetry by setting
\bea
m(r,t)&=&M(t)r^3,\label{eq:flatmass}\\
R(r,t)&=&\frac{S(t)}{S_1}r\equiv \tilde S(t)r,\label{eq:scale}
\eea 
so that the Dimensions of $M$ are those of $\rho$. The scale factor $S$ is Dimensionless, but it is convenient to involve an arbitrary value at the time of the symmetry breaking instability $t_1$ . These expressions for the mass  and circumferential radius guarantee the geometric flatness of the model. 

Moreover the energy density is taken to be 
\be
\rho=\rho_V(t)+\rho_\nu(t)+\rho_d(t),\label{eq:totenergydensity}
\ee
where $\rho_V$ is the vacuum energy density, $\rho_\nu$ is the  zero rest mass  (effectively or really; including neutrinos) energy density, and $\rho_d$ is the zero pressure matter density, which we assume to include the dark and baryonic  matter. We normally write
\be
\rho_m\equiv \rho_\nu+\rho_d.\label{eq:matterdensity}
\ee

The cosmological pressure  at any epoch is therefore
\be
p=\frac{\rho_\nu}{3}-\rho_V.\label{eq:pressure}
\ee
The vacuum energy is taken to be
\be
\rho_V= \frac{c^4\Lambda(t)}{8\pi G}, ~~~~~~\Lambda(t)=\frac{\lambda}{c^2t^2},\label{eq:vac}
\ee
where we have restored the units temporarily to indicate that $\lambda$ is a number.. 

We may write the equation of state of the matter as 
\be
w(\tilde S)=\frac{p_m}{\rho_m}\equiv \frac{\rho_\nu/3}{\rho_m}\equiv \frac{1}{3(1+\frac{\rho_d(1)}{\rho_\nu(1)}\tilde S)},\label{eq:EOS}
\ee
where $(1)$ indicates the value at $\tilde S=1$, if the epoch of symmetry breaking is $t_1$. This time is a parameter that we use as a unit of time subsequenly.  We have used the usual scale dependences of the `photon'  density and the dust density that follow from equation (\ref{eq:massevolution}). 
 
The initial ratio of the `dust' density to the `zero rest mass' density appears as another parameter in the problem.   However the initial value is related to the value at any epoch by
\be
\frac{\rho_d(1)}{\rho_\nu(1)}=\frac{1}{\tilde S}\frac{\rho_d}{\rho_\nu}.\label{eq:kdevolution}
\ee
Applying this at the current epoch and at equipartition we infer that 
\be
\frac{\rho_d(1)}{\rho_\nu(1)}=\frac{z_{eq}}{\tilde S_o}, \label{eq:dnuratio}
\ee
which is very small. The value of $z_{eq}$ according to (\cite{Planck2018}) is $3387$. 

Given $z_{eq}$,there is now an iterative procedure necessary because before integration of the governing equations  (see equations \ref{eq:work1} and \ref{eq:work2}) we must assign  an assumed $\tilde S_o$.  The iteration is continued until the integration arrives at our epoch (defined by an acceptable Hubble-Lemaitre value coincident with an acceptable ratio of matter to vacuum density) with the starting value of  $\tilde S_o$.  This procedure defines the  global quantities of the model as summarized below.




We continue formally by substituting the form for $m$ and $R$ from equations (\ref{eq:flatmass}) and (\ref{eq:scale}) into equation (\ref{eq:mass})  to find 
\be
M(t)=\frac{4\pi}{3}\tilde S^3\rho_V(1+\Omega_{mv}),\label{eq:scalemass}
\ee
where we have  again set  $\tilde S\equiv S/S_1$ and 
\be
\Omega_{mv}\equiv \frac{\rho_m}{\rho_V}.\label{eq:mass/vac}
\ee
These are key variables in our analysis.

We can now substitute for the mass from equations (\ref{eq:scalemass}) and (\ref{eq:flatmass}) into equation (\ref{eq:energyinit}) together with the scaled form of $R$. This requires  us to split the equation into two equations
\bea
e^\omega&=&\tilde S^2,\label{eq:metricoefficient}\\
\frac{\dot S^2}{S^2}&=&\frac{8\pi}{3}\rho_V(1+\Omega_{mv}).\label{eq:Hsquared}
\eea
Equation (\ref{eq:metricoefficient}) confirms the flatness of the metric. Recalling equation (\ref{eq:vac}) we write equation (\ref{eq:Hsquared}) in the useful form
\be
t~h=\sqrt{\frac{\lambda}{3}}\sqrt{1+\Omega_{mv}}\equiv t\frac{\dot S}{S}\equiv t\frac{\dot{\tilde S}}{\tilde S},\label{eq:work1}
\ee
which defines the Hubble variable $h(t)$. 

Equation (\ref{eq:energycons}) can now be written explicitly as an equation for $\Omega_{mv}(t)$ using equation (\ref{eq:work1}) as 
\be
t~\dot\Omega_{mv}=2(1+\Omega_{mv})-\sqrt{3\lambda}(1+w(\tilde S))\Omega_{mv}\sqrt{1+\Omega_{mv}},\label{eq:work2}
\ee
which must be solved together with equation (\ref{eq:work1}).

Once solved, these two equations define a cosmology with some interesting aspects. The iterative procedure provides the uniqueness of the model. to emphasize this we  give the  iterative results here which are to be used in additional discussion below. 

The iteration yields $\rho_d(1)/\rho_\nu(1)=2.512\times 10^{-33}$ with an initial $\tilde S_o=1.348\times 10^{36}$ yielding after iteration the  corresponding value $\tilde S_o=1.347\times 10^{36}$. This occurs at $\ln{t_o/t_1}=41.324$, which gives the age of our epoch  as $28.0 Gyr$ if $t_1$ is chosen to be $1s$. This agrees with a direct calculation from equation (\ref{eq:work1}) below. 
the value of $\lambda$ is given as $\sqrt{3\lambda}=5.2134$ and $\Omega_{mv}(0)=0.4641$. These quantities require  a Hubble/Lemaitre value at the current epoch of $73.36 km/s/Mpc$ and a current $\Lambda=\lambda/(c^2t_o^2)$ equal to  $1.29\times 10^{-56} cm^{-2}$ all assuming $t_1=1s$. 

These values are remarkably close to concordance values with the exception of the factor two in the age of the Universe. we discuss this further in the next section. It should be emphasized that because inflation is absent in this model, the initial vacuum fluctuations may not be as in the concordance model. However all of the relevant physics and analysis leading to the explanation  of structure arising  since the equipartition epoch is applicable to this model. Only the extra time available is a major difference. There is a  very small creation of matter from the vacuum at any epoch where $\Omega_{mv}$ has not attained its ultimate value (i.e. $\tilde S\ne\infty$).


\section{Cosmological model details }
\label{sect:cosmodstruct}

\begin{figure}{} 
\begin{tabular}{cc} 
\rotatebox{0}{\scalebox{0.4} 
{\includegraphics{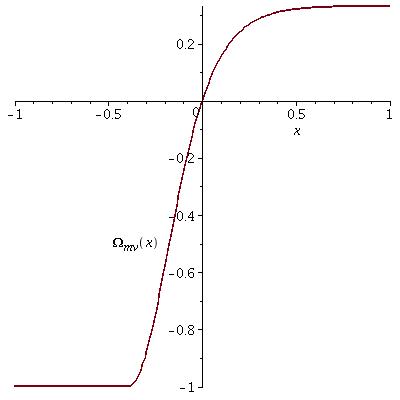}}}
{\rotatebox{0}{\scalebox{0.4} 
{\includegraphics{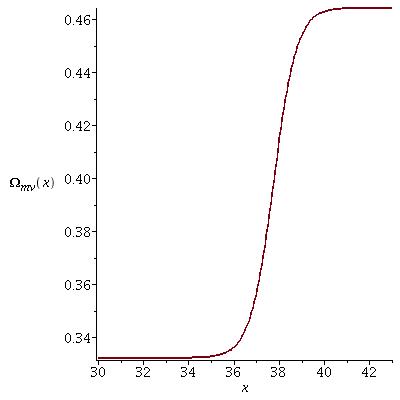}}}}
\end{tabular}
\caption{ The figure  on the left shows the beginning of this Universe model as an instability in a zero energy vacuum that causes $\Omega_{mv}$ to transition rapidly from $-1$ to an early asymptotic value $\sim 0.332$.  This rises rapidly again  at a subsequent epoch of rapid matter creation beginning near $x=35$, as shown on the right panel. The value of $\sqrt{3\lambda}\approx 5.2134$ in order to give $\Omega_{mv}(o)\equiv \rho_m(o)/\rho_V(o)\approx 0.4642$. When $\Omega_{mv}=-1$ the total energy is zero.  The figure on the right shows the delayed transition. On the left the vacuum $\tilde S\approx 0.663$ and on the right $\tilde S\approx 2.951\times 10^{30}$.}
\label{fig:omegamv}
\end{figure}

In figure (\ref{fig:omegamv}) we show $\Omega_{mv}(x)$ where $x=\ln(t)$.  At $x=0$  (or $t/t_1=1$) we have automatically $\tilde S(0)=1$. The unit of time $t_1$ is also a free parameter, but we have agreement with current global values when $t_1=1s$. We have iterated $\lambda$ and $\tilde S_o$ until the asymptotic value $0.4641$ at large $x$ is nearly that given by Planck satellite measurements  (\cite{Planck2018}; see also \cite{AKW2020}, where the number is approximately $0.461=0.316/0.685$).  A good value of the asymptotic  $\Omega_{mv}$ at large enough $\tilde S$ is found  from equation (\ref{eq:work2} by setting $w=0$ as
\be
\Omega_{mv}(\infty)=\frac{2}{k^2}(1+\sqrt{1+k^2}),\label{eq:omegamvinfty}
\ee
and hence 
\be
k^2=4\frac{1+\Omega_{mv}(\infty)}{\Omega_{mv}(\infty)^2}.\label{eq:kinfinity}
\ee
We have set $k=\sqrt{3\lambda}$.

The important behavioural feature shown in the left panel of figure (\ref{fig:omegamv}) is the abrupt rise of $\Omega_{mv}$ from the early zero energy value $-1$ (the classical vacuum) at $x<0$, to the first asymptotic value equal to $\sim 0.332$ at $x\ge1$. This is the symmetry breaking instability (any variation of $\Omega_{mv} >-1$ initiates the instability according to equation (\ref{eq:work2})) that we attribute to the electroweak symmetry breaking.  

The right hand panel of the figure shows the subsequent rise to the current value $\sim 0.4642$. This rise is not quite so abrupt as the initial instability. It corresponds to the equation of state (EOS)  declining towards zero from the massless EOS value of $1/3$.  Equation (\ref{eq:work2}) shows that this leads to a rapid increase in $\Omega_{mv}$.

Throughout this note we refer to the usual cosmic co-moving  flat FLRW  coordinate time scale, although we calculate the logarithm.  As many have recognized the logarithmic time $x$ has a great advantage in recording the history of the Universe. In these terms the zero energy vacuum state has persisted for an infinity, before the world creating instability. This removes awkward questions about `before the beginning',  if indeed the Universe unfolds in logarithmic time. This is the case for equations (\ref{eq:work1}) and (\ref{eq:work2}) that define our model.

\begin{figure}{}
\begin{tabular}{cc} 
\rotatebox{0}{\scalebox{0.4} 
{\includegraphics{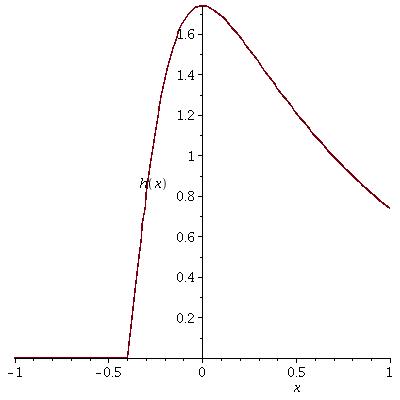}}}&
\rotatebox{0}{\scalebox{0.4} 
{\includegraphics{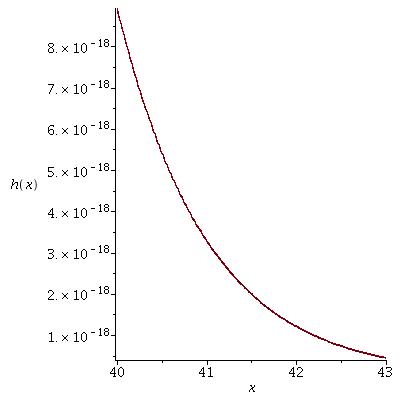}}}\\
{\rotatebox{0}{\scalebox{0.4} 
{\includegraphics{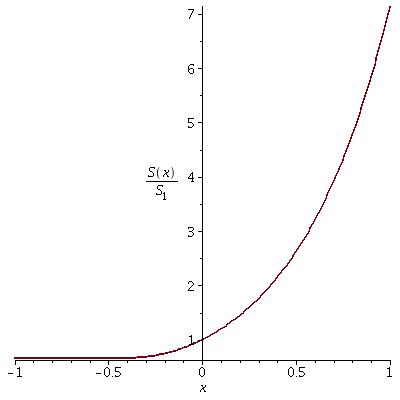}}}}&
\rotatebox{0}{\scalebox{0.4} 
{\includegraphics{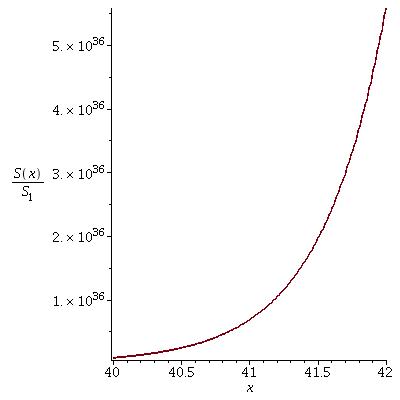}}}
\end{tabular}
\caption{In the upper row we show the Hubble/Lema\^itre variable as a function of $x$ near the initial instability on the left and approaching the current epoch on the right. The only era in which $h$ actually increases is during the instabiity. The lower row shows the corresponding scale factor near the transition epoch on the right and much later near the current epoch on the right.  The initial variation is $S\propto t^{2.0}$ for $x<\sim 35$  on the left while in the asymptotic region on the right panel $S\propto t^{2.1}$.   }
\label{fig:scaling}
\end{figure}

The expansion of the Universe is indicated in the lower panels of figure (\ref{fig:scaling}). It begins in the lower left panel with the sudden predominance of the vacuum energy when $\Omega_{mv}$ fluctuates upwards from $-1$.  Using our parameters in equation (\ref{eq:work1}) we find that $\tilde S\propto t^{2.0}$ in the early asymptotic stage on the lower left panel, and $\tilde S\propto t^{2.1}$ in the late asymptotic stage on the lower right panel. These correspond to classical acceleration parameter $q=-\ddot S/(h^2 S)$ of $q=-0.5$ and $q=-0.524$ respectively. The latter value is the value at the current epoch, which would be $-1$ for exponential expansion.

The upper row  in figure (\ref{fig:scaling}) shows the Hubble/Lemaitre variable  at the initial instability on the left and near the current epoch on the right.
The log time at the current epoch as a result of the iterations is $x_o=41.324$, from which all of our values follow.


We have assumed that the  vacuum variation with $1/t^2$  begins with the initial instability and continues driving the subsequent accelerating expansion. We require in effect the entire Universe to behave as a coherent excitation, namely  as a long wavelength, scalar, pseudo Goldstone (Higgs) Boson particle. This is hypothetically coincident with the electroweak symmetry breaking. The wavelength of this `particle'  is $1/\sqrt{\Lambda}$, which is equal approximately to the apparent horizon, given  the inverse square temporal variation of $\Lambda$.

After our iteration to obtain a self consistent $\tilde S_o$ and a current value of $\Omega_{mv}$  we find $t_oh_o$ from equation (\ref{eq:work1})  and from the direct integration. The iterated values of $\lambda$ and $\Omega_{mv}(o)$ are used on the RHS of equation (\ref{eq:work1}). The resulting self consistent $h_o$ and $t_o$  and $\Lambda_o=\lambda/t_o^2$ are, as stated previously,
\be
t_o=28.0 Gyr,~~~~~~~~~~h_o=73.36 km/s/Mpc,~~~~~~~~~~\Lambda_o=1.29\times 10^{-56} cm^{-2}.\label{eq:keyresults}
\ee

Given $z_{eq}=3387$ the coordinate time $t_{eq}\approx (1/z_{eq})^{1/2.1}t_o\approx 0.58 Gyr$. Thus there is a rather longer time in which perturbations above the Jeans scale can grow into observed structure. The linear growth rate using the standard formula (e.g.\cite{P1999})  is  however reduced to $t^{\sim 0.57}$ compared to $t^{2/3}$ for a matter dominated Universe. We have used $S\propto t^{2.1}$.

It may be nevertheless that primordial black holes and/or galaxies will have appeared  quite soon after the recombination ($\sim 1.0Gyr$), having grown linearly since $z_{eq}$, which is already an interval of $0.42Gyr$. That is, the extra time may allow perturbations to go non linear and detach from the expansion `earlier' relative to us. Abundant such objects at high red shifts would be one way of distinguishing this model from the concordance model. The counter tendency is the rapid historical expansion rate. This may  lower the  physical density of the detached objects.

 
 
 
 The rapid expansion rate removes the horizon problem for homogeneity just as does inflation. If $\tilde S\propto t^2$ as in the earlier asymptote, then the particle horizon size at time $t >t_{ph}\ge t_1$ can be written as (taking $\tilde S_{ph}=1$)
 \be
 R_{ph}(t)=ct^2(\frac{1}{t_{ph}}-\frac{1}{t}).\label{eq:PH}
 \ee
Here $t_{ph}\ne 0 $ is the earliest time from which light signals arrive at $t$. This implies that a causally connected patch expands faster than the apparent horizon of the Universe.  However, in our model the vacuum metastable state is already homogeneous (but for quantum fluctuations) whicheffectively removes the problem.




During the  zero energy vacuum era the entropy is essentially zero by the third law of thermodynamics (plus no local gravitational degrees of freedom (cf \cite{Pen2004},\cite{Pen2010}) and the model is homogeneous, although presumably subject to quantum noise. This  initial condition  avoids two of the criticisms of the big bang model, without inflation.  Bouncing cosmological models encounter the black hole entropy problem (\cite{Pen2004}), which has driven a conformally based cycling model \cite{Pen2010}.

 At equipartion  $\tilde S_{eq}\sim 2.3\times 10^{34}$. This implies formally  an enormous temperature, exceeding the Planck temperature $kT_P/=m_Pc^2$, (or $T_P\sim 1.4\times 10^{32}K$) at $\tilde S_1=1$. This does not correspond to the expected temperature of symmetry breaking,  $T_1=159.5\pm 1.5$ GeV, that is $T_1\sim 1.85\times 10^{15}K$.  This would place us in the region of Grand Unified Theory (GUT) symmetry breaking  (assuming the temperature can not in fact exceed the Planck value) about which very little is certain.
 
\begin{figure}{}
\begin{tabular}{cc} 
\rotatebox{0}{\scalebox{0.4} 
{\includegraphics{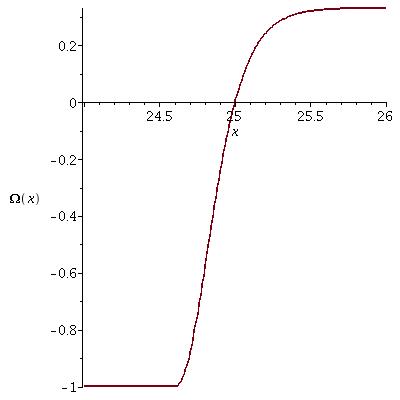}}}&
\rotatebox{0}{\scalebox{0.4} 
{\includegraphics{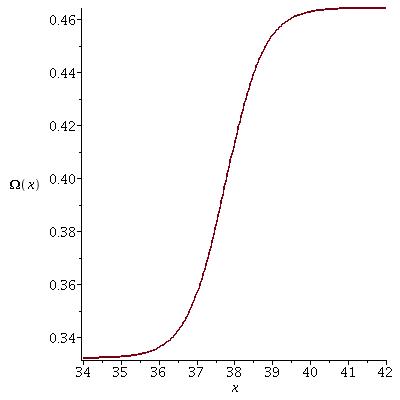}}}\\
{\rotatebox{0}{\scalebox{0.4} 
{\includegraphics{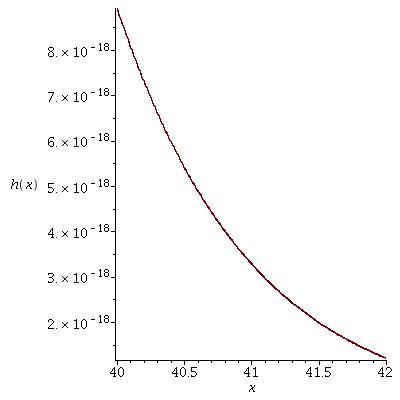}}}}&
\rotatebox{0}{\scalebox{0.4} 
{\includegraphics{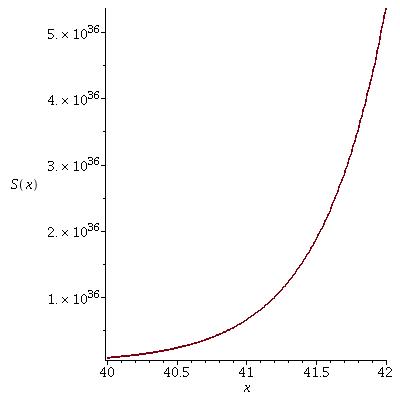}}}
\end{tabular}
\caption{In the upper row we show on the left $\Omega_{mv}$  as a function of $x$ near the delayed instability, and the `equation of state' driven transition to the current epoch value on the right panel. The parameter $\lambda$ is the same as in figure (\ref{fig:omegamv}). The lower row shows the Hubble/Lemaitre variable near the current epoch on the left panel and the scale variable near  the current epoch on the right. }
\label{fig:Sscaling}
\end{figure}

However it is possible to delay the expansion in order to fit an electro weak transition in the following fashion. 
We set $\Omega_{mv}(x=25)=0$ so that the vacuum state continues until $t=e^{25}s\sim 2.28\times 10^3 yrs$. At this value of $ x$ we give $\tilde S$ the same value that it would have in figure (\ref{fig:scaling}) namely $\sim 5.745 \times 10^{21}$.  This delayed model is shown in figure  (\ref{fig:Sscaling}).

There is no essential change in our values for $t_o$, $h_o$ and $\Lambda_o$. We find, due mainly to the different integration interval, that the self consistent $\tilde S_o=1.295\times 10^{36}$, rather than the previous $1.348 \times 10^{36}$. We observe in  the upper row of figure (\ref{fig:Sscaling}) the early and late transitions in $\Omega_{mv}$, while the lower row shows the current behaviour of $h(x)$ and $\tilde S(x0$. 

The value $\tilde S(25)=5.745\times 10^{21}$. This implies a temperature ratio relative to the recombination epoch ($z_{rec}=1087$) of $\tilde S(rec)/\tilde S(25)=1.191\times 10^{33}/5.745\times 10^{21}$. This yields $T(25)/T_{rec}=2.073\times 10^{11}$. The temperature $T_{rec}$ is consistently $\sim 3000K$ if $T\propto 1/\tilde S$, whence $T(25)=6.227\times 10^{14}K$.  This is within a factor $3$ of the electroweak transition temperature. If we use the scale factor inside the vacuum of $\tilde S_{vac}=3.811\times 10^{21}$, then $T(vac)=9.34\times 10^{14}$, a factor of $2$ low. It seems that a slight adjustment  of the instability to slightly earlier times could remove even this discrepancy.   

This resolves the temperature problem at the presumed elctroweak symmetry breaking.  Other quantities remain essentially the same and the age age of the Universe remains the outstanding disagreement with the concordance model.

\section{Discussion}
Our fundamental assumption is that the dark energy evolves rather slowly ($\rho_V\propto 1/t^2$) compared to the inflationary era. We have shown that the best justification of this assumption is through Dimensional analysis. The assumption also ensures that the vacuum scale is always similar to the apparent horizon of the Universe, which suggests that he Universe is a long wave length pseudo-Goldstone excitation.

Our model requires an older Universe than that of the concordance model. The material Universe is due to an instability  of a  zero energy vacuum state which we take to coincide  with  electroweak spontaneous symmetry breaking. The nature of matter and  the positive energy vacuum (`dark energy')  in the Universe and their relative  importance therefore follows from this event. That is, we assume that the electroweak symmetry breaking  produces immediately the matter to vacuum ratio $\rho_m/\rho_v\equiv \Omega_{mv}$. 

We expect therefore; that the decay products of the $W^{\pm}$ and  $Z^0$ vector bosons and their subsequent interactions, plus high energy photons and their interactions, to comprise both the visible and dark  material Universe plus the  `Dark Energy' ($\rho_V$). The Higgs Boson is essential to the symmetry breaking of the gauge invariance, and its decay products will also be present. The details of this process are certainly beyond the scope of this article, but there is a peculiar coincidence that should be noted. 

The Weinberg or Weak Mixing Angle  $\theta_W$ is given in the most recent CoData compilation as $\sin^2{\theta_W}=0.22290$, which is $\sin{\theta_W}= 0.472$.  This is at an energy of $91.2 GeV$. The angle is not strictly constant appearing to `run' with energy. For example at $7~ TeV$ the ATLAS collaboration at the LHC gives $sin^2(\theta_W)=0.23080\pm 0.00120$ or $\sin{\theta_W}=0.479$. Proceeding with the lower value (closer to the electroweak energy of symmetry breaking), we have according to the theory,
\be
\sin{\theta_W}=\frac{\sqrt{M_Z^2-M_W^2}}{M_Z}\approx 0.472,
\ee 
where $M_Z\approx 91.2 GeV$ and $M_W=80.38 GeV$ are the corresponding boson masses.

To within errors, this is almost the value of  the current epoch matter to `dark energy' ratio $\rho_m/\rho_V\approx 0.315/0.685=0.460$ ($0.32/0.68=0.4705$). The current number is a little smaller than what we have used for our calculations  namely $0.4642$, although that value is within observational divergences and can be refined. This suggests that difference in the decay chains of the bosons normalized by the decay of the $Z^0$ boson, yields the matter to dark energy ratio in the Universe. 

Note that the matter includes the dark matter, so the dark matter must be present in the difference between the $Z^0$ and $W^{\pm}$ evolutionary sequences.  Perhaps one way that the matter antimatter symmetry inherent in the bosons can be broken is if the $Z^0$ decay chain interacts asymmetrically with the $W$ decay chain. The different masses of the bosons may lead to differences in equilibrium abundance of the decay products. The dark energy by contrast appears to be due solely to the decay chain beginning with the production of the $Z^0$ bosons. 


Astrophysically, the dominant change in this model is the age of the Universe. This may allow structure to grow at higher redshifts then currently anticipated. Searches for fully formed galaxies or Quasars at ever higher red shifts might provide a decisive test of this model. Philosophically, the initial zero energy vacuum appears as the matter/antimatter balance that yields nothing. Fortunately it is unstable.

\section{Acknowledgements}

I thank Judith Irwin for her constant encouragement, criticism and help.

\section{Appendix A:Cosmological Numbers}

In \cite{HIr2019}) it is argued that the Buckingham theorem confirms that statements about the Universe should amount to  expressions between Dimensionless `numbers'. With $n=9$ relevant quantities ($h,\bar\rho,\Lambda,T_b,\delta T_b,R_{AH},t,c,G$) and $m=4$ independent Dimensions (length, time, mass, temperature) for the homogeneous classical  Universe we expect a basic function of five ($n-m$; e.g. \cite{Hen2015}) variables. We omit the Mond possibility here, but see \cite{HIr2019}.  We have written $R_{AH}$ for the apparent horizon and $T_b$, $\delta T_b$ for the background temperature and temperature fluctuation.

The most general expression involving all independent `cosmic numbers' (not all need ultimately appear) would be some single continuous  `cosmic' function
\be
f(K,V,F,T,R)=0,\label{eq:Cosfct}
\ee
where a complete (but not unique) set of `numbers' at some epoch follow from our cosmic catalogue as ($t$ is the epoch age)
\bea
K&\equiv& \frac{8\pi G\bar\rho}{3h^2},\label{eq:K}\\
V&\equiv& \frac{\Lambda c^2}{3h^2},\label{eq:V}\\
F&\equiv& \frac{\delta T_b}{T_b},\label{eq:F}\\
T&\equiv& ht,\label{eq:T}\\
 R&\equiv&\frac{R_{AH}}{ct}.\label{eq:RAH}
\eea

All of these numbers are of $O(1)$ \footnote{$K$ is $\Omega_m$ but the numerical factor of 3 and the factor $8\pi$, which factors can not be realized entirely by Dimensional argument} according to currently accepted values, except the number $F$. It is of $O(10^{-5})$  and can be eliminated from the relation (\ref{eq:Cosfct}) by a zeroth order MacLaurin expansion in $F$. It is the only number that corresponds to a definite epoch and certainly contains information about the history of the Universe. However the MacLaurin expansion suggests that it does not affect directly the global properties of later epochs (ignoring sub structure of the homogeneous Universe).  Moreover the apparent horizon of  a flat Universe is essentially $c/h$  (e.g. \cite{Rin2006}) so that $R$ is not  independent of $T$. This leaves the numbers  $K$, $V$ $T$ as arguments  of the cosmic function, which each contain the Hubble variable $h(t)$ so that the cosmic function becomes 
\be
\tilde f(K,V,T)=0.\label{eq:cosfct2}
\ee

We note that  the fractional expansion rate $h$ varies as  $1/t$ in most cosmological models (including this one and excepting  de Sitter). This makes $T$  and normally $K$ `universal' constants. In order to avoid making our epoch special we need only assume that $V$ is also constant. In fact on general grounds, we expect  the numbers $K$, $V$ and $T$ to  be close to unity at all epochs.  Otherwise very big  numbers (or rather their reciprocals) or very small numbers, would be eliminated as was the number $F$. 
 
This argument requires $h\propto 1/t$, and $\bar\rho\propto 1/t^2$ as is known,  but also $\Lambda\propto 1/t^2\propto \bar\rho$. The counter argument to this conclusion is based on the anthropic principle, but this is not readily testable. In principle the values of the numbers at different epochs are measurable.

We have used only the Buckingham theorem, a reasonable choice of the set of Universe characteristics (a `catalogue' of properties), and a conviction that important numbers should be $O(1)$, in order to reach the form of the cosmic function. We have also assumed that our epoch is not unique. There is no use of dynamics to this point.

However  as an example, the Friedman equation in  the concordance model gives with zero curvature
\be
h\equiv \sqrt{\frac{8\pi G}{3}\bar\rho+\frac{\Lambda c^2}{3}},
\ee
which can be written as ($h\ne 0$) the familiar 
\be
1-(V+K)=0.\label{eq:fexamp}
\ee
 This indicates a degeneracy between the numbers $V$ and $K$ when $T$ is non zero, and the Universe is spatially flat. Agreement with the general form  (\ref{eq:cosfct2}) follows trivially by multiplying equation(\ref{eq:fexamp}) by T. One does not normally resolve the function (\ref{eq:cosfct2}) without additional information, in this case General Relativity. However once one assumes that $T$ is constant, the degeneracy between $K$ and $V$ is inevitable although its form is not.
 In the concordance model $K=0.32$ and $V=0.68$ although more recently (\cite{AKW2020}, \cite{Planck2018}) these are $K=0.315$ and $V=0.685$.

\


\begin{thebibliography}{}
\bibitem[Abdullah, Klypin\& Wilson(2020)]{AKW2020}Abdullah, M.H., Klypin, A. \& Wilson, Gillian, Astrophysical Journal, {\bf 901}, \# 2, 2020
\bibitem[Berestetskii, Lifshitz and Pitaevskii(1971)]{BLP1971} Berestetskii, V.B., Lifshitz, E.M.,and Pitaevskii, L.P., "Relativistic Quantum Field Theory", Pergamon (Addison-Wesley), Oxford, {\bf p36}, 1971
\bibitem[Bicknell\& Henriksen(1979)]{BH1979} Bicknell, G.V. \& Henriksen, R.N., Astrophysical Journal, {\bf 232}, 670, 1979
\bibitem[Henriksen(1982)]{Hen1982} Henriksen, R.N., Physics Letters B, {\bf 119}, 85, 1982
\bibitem[Henriksen, Emslie and Wesson(1983)]{HEW1983} Henriksen, R.N., Emslie, G.A. and Wesson, P.S., Phys. Rev.D, {\bf 27}, 1219, 1983
\bibitem[Henriksen(2015)]{Hen2015} Henriksen, R.N., "Scale Invariance",Wiley-VCH, Weinheim, Germany, 2015
\bibitem[Henriksen and Irwin(2019)]{HIr2019} Henriksen, R.N. and Irwin, J.A., "A `Numbers 'Approach to Astronomical Correlation",ArXiv190208704, 2019
\bibitem[Lykken \& Spiropulu(2013)]{LS2013} Lykken, J. \& Spiropulu, Maria, Physics Today, {\bf 66}, \# 12, 28, 2013
\bibitem[Misner(1969)]{Mis1969}Misner, C.W., in "Astrophysics and General Relativity", v1,Chr\'etien,M.,Deser, S. and Goldstein, J. (eds), Gordon and Breach, London, chpt 3, 1969
\bibitem[Oesch,Brammer, van Dokkum et al.(2016)]{OBD2016} Oesch, P.A., Brammer,G., van Dokkum, P.G. and 15 others, Ap.J.. {\bf 819}, 129,2016
\bibitem[Olson and Jordan(1987)]{OJ1987} Olson, T. and Jordan, T., Phys. Rev. D, {\bf 35}, 3258, 1987
\bibitem[Ostriker and Steinhardt(1995)]{OS1995} Ostriker, J.P.  and Steinhardt, P., Nature, {\bf 377}, 600, 1995
\bibitem[Peacock(1999)]{P1999} Peacock, J. A., "Cosmological Physics", Cambridge University Press, Cambridge, UK, 1999
\bibitem[Penrose(2004)]{Pen2004} Penrose, R., "The Road to Reality", BCA(Jonathan Cape), G.B., chpt. 27, 2004
\bibitem[Penrose(2010)]{Pen2010} Penrose, R., "Cycles of Time", The Bodley Head , London, 2010
\bibitem[Peter(2013)]{Peter2013} Peter, P., arXiv1303.2509, 2013
\bibitem[Planck Collaboration(2020)]{Planck2018} Planck Collaboration, Astron. and Astrophys., {\bf 641}, A6, 2020
\bibitem[Ratra and Peebles(1988)]{RP1988} Ratra, B. and Peebles, P.J.E., Phys. Rev. D, {\bf 37},3406,1988
\bibitem[Rindler(2006)]{Rin2006} Rindler, W.,"Relativity". second Ed., Oxford University Press, New York, chpt. 17, 2006
\end{thebibliography}
\end{document}